\documentclass[iop,12pt]{emulateapj}
\usepackage{natbib} 
\usepackage{amssymb,amsmath}
\usepackage[dvipsnames]{color}
\def\refnew#1{\,(\ref{#1})}

\def\kc{{k_c}}
\def\H{\mathcal{H}}
\def\taun{\tau_n}
\def\taus{\tau_s}
\def\Gyr{\,\mathrm{Gyr}}
\def\dynpcmsq{\,\mathrm{dyn\, cm^{-2}}}
\def\gpcc{\,\mathrm{g\, cm^{-3}}}
\def\Asep{A_{\mathrm{sep}}}
\def\Aout{A_{\mathrm{out}}}
\def\Ain{A_{\mathrm{in}}}
\def\Ac{A_c}
\def\kcap{k_{\mathrm{cap}}}
\def\Hsep{\mathcal{H}_{\mathrm{sep}}}

\def\phimax{\phi_{\mathrm{max}}}

\def\km{\,\mathrm{km}}
\def\Myr{\,\mathrm{Myr}}
\def\R{\mathcal{R}}

\begin{document}

\title{Special mean motion resonance pairs: Mimas-Tethys and Titan-Hyperion}
\author{Jing Luan}
\email{jingluan@caltech.edu}
\affiliation{California Institute of Technology, Pasadena, CA 91125, US}

{\large \begin{abstract}

Five pairs of large solar system satellites occupy first order mean-motion resonances (MMRs).  Among these, the pairs of Mimas-Tethys and Titan-Hyperion are special.  They are located much deeper in resonance than the others and their critical arguments librate with much greater amplitudes.  These characteristics are traced to the insignificant damping, over $\Gyr$ timescales, of Mimas's orbital inclination and Hyperion's orbital eccentricity.  Absent that, these resonances would not survive.  Instead their librations would be overstable and escape from resonance would occur on the relevant damping time.  Unlike the aforementioned MMRs, those involving Enceladus-Dione, Io-Europa, and Europa-Ganymede are limited by eccentricity damping.  
They must either remain at the shallow depths they currently occupy, or, if they venture deeper, retreat after a limited time. The latter seems almost certain for Enceladus-Dione and quite likely for the others,  We examine the MMRs involving Mimas-Tethys and Titan-Hyperion under the assumption that they formed as a result of convergent migration.  Capture probabilities are $\sim 6\%$ for the former and $100\%$ for the latter.  The possibility of collisional excitation of their large librations is investigated but largely discounted.

\end{abstract}

\maketitle

\section{Introduction}\label{sec:intro}
Several pairs of solar system satellites are involved in mean motion resonances (MMR)  \citep{1976ARA&A..14..215P}. \cite{1965MNRAS.130..159G} proposed that most of them formed as the satellites' orbits expanded due to tidal torques from their parent planets. Observations of volcanoes on Io and geysers on Enceladus suggest that significant tidal evolution is ongoing. Relative to the others, the MMRs between Mimas-Tethys and Titan-Hyperion stand out in two respects.  They are located much deeper in resonance\footnote{Depth in resonance is reckoned  relative to that at which a separatrix first appears in the 2D phase space.} and their critical arguments librate with much larger amplitudes.  Our investigation explains each of these special features.

Unless halted by some dissipative process, convergent migration inexorably drives resonant satellite pairs deeper into MMR.  Damping of the orbital inclination of Mimas and the orbital eccentricity of Hyperion would be the relevant dissipative processes for the Mimas-Tethys and Titan-Hyperion MMRs, but each is negligible. This accounts for their large resonance depths. It also allows us to trace the libration amplitudes back to when the resonances formed.    

Dissipation associated with tides raised in Saturn by Mimas causes Mimas to migrate toward Tethys, making a tidal origin for this MMR plausible. Titan is so far away from Saturn that it migrates at a negligible rate.  \cite{1978Icar...36..240P} hypothesized that multiple bodies formed within the MMR with Titan and that Hyperion is the only one to survive. However, particle eccentricities excited within a MMR depend sensitively on semi-major axis implying that impacts between particles would occur at high relative velocities, a situation that is not conducive to accretion.  Thus we favor the idea that the Titan-Hyperion MMR arose by convergent migration, although the responsible mechanism is uncertain. 

Previous investigations of the Mimas-Tethys system by \cite{1969AJ.....74..497A} and \cite{1972MNRAS.160..169S} suggested that its capture in MMR was a low-probability event. In arriving at this conclusion, these authors applied an approximate method based on a perturbed pendulum to estimate capture probabilities. We take advantage of more accurate capture probabilities originally derived by \cite{1979CeMec..19....3Y}, then elucidated by \cite{1982CeMec..27....3H} and finally presented in simple form by \cite{1984CeMec..32..127B}.  Our results generally support the findings of \cite{1969AJ.....74..497A} and \cite{1972MNRAS.160..169S}. 

We treat the less massive satellite in each pair as a test particle. Hamiltonian dynamics near MMRs of interest to us is dominated by terms with a single resonance argument, $\phi$. This eliminates one degree of freedom and gives rise to a constant of motion, denoted here by $k$, which is the term of leading order involving the perturber's mass in the Jacobi constant. The next order term provides an independent constant of motion, $\H$. The use of $\H$ is appropriate because it serves as the Hamiltonian in MMR dynamics. Higher order terms are combinations of $k$ and $\H$, and thus do not introduce additional independent constants of motion. For Mimas-Tethys, the canonical conjugate position and momentum are $\phi$ and $s^2$, where $s\equiv \sin(I/2)$ with $I$ Mimas's orbital inclination with respect to Saturn's equator plane. For Titan-Hyperion, they are $\phi$ and $e'^2$, where $e'$ is Hyperion's orbital eccentricity. We employ unprimed and primed parameters for the inner and outer satellites.

The topology of the contours of constant $\H$ is determined by $k$. An unstable saddle point appears for $k>\kc$. Both $k$ and $\H$ evolve due to migration and dissipation, but far slower than the quasi-periodic motion along contours of constant-$\H$. In the main text, we list formulas for $k$ and $\H$ for each MMR.   We use the same symbols although their definitions vary for different MMRs.

Our paper is constructed as follows. Section \ref{sec:M-T} analyzes the Mimas-Tethys MMR. We demonstrate that impacts are inadequate to produce its libration. Titan-Hyperion is discussed in Section \ref{sec:T-H}. Finally, we conclude in Section \ref{sec:conclusion}.  

\section{Mimas-Tethys}\label{sec:M-T}
Tethys is as ten times massive as Mimas, so we treat the latter as a test particle.
There are three inclination MMRs with period ratio near $2:1$\footnote{These are more properly classified as $4:2$ resonances.} with disturbing function, $\R$, and associated resonant argument, $\phi$, given by $\R\propto s^2$ with $\phi=4\lambda'-2\lambda-2\Omega$,  $\R\propto ss'$ with $\phi=4\lambda'-2\lambda-\Omega-\Omega'$, and $\R\propto s'^2$ with $\phi=4\lambda'-2\lambda-2\Omega'$. Here $\Omega$ is the ascending node, and $\lambda$ is the mean longitude. Nodal precession due to Saturn's oblateness separates them spatially so each resonance can be analyzed separately.  Mimas-Tethys occupy the mixed-$ss'$ resonance.  As Mimas migrated toward Tethys, it almost certainly encountered the $s^2$ MMR first but avoided capture. Shortly thereafter it was captured in the $ss'$ MMR where it is likely to remain and thus never reach the $s'^2$ MMR. Next we estimate the capture probability for the $s^2$ MMR.

\subsection{$s^2$ MMR}
The associated $k$ and $\H$ take the forms
\begin{eqnarray}
k&=&6 s^2+{1\over 2}\kc\cos\phi+\frac{\dot \phi}{2n}\, ,\\
\H&=& k s^2-3 s^4-s^2 \kc\cos\phi\, ,
\end{eqnarray}
where $\phi=4\lambda'-2\lambda-2\Omega$. The critical $k$ reads
\begin{equation}
\kc=\mu'\frac{f_{57}}{2}\frac{a}{a'}\simeq 2.81\times 10^{-7}\, .
\end{equation}
Here $\mu'\simeq 1.09\times 10^{-6}$ is the mass ratio of Tethys to Saturn and $f_{57}\simeq 0.82$. Orbital semi-major axes for Mimas and Tethys are $a\simeq 1.8552\times 10^5\km$ and $a'\simeq 2.9466\times 10^5\km$ \citep{1999ssd..book.....M}. The topology of the phase space at $k>\kc$ is depicted in Fig.\refnew{Contours-s2}. There are three fixed points at which both $\partial\H/\partial\phi$ and $\partial\H/\partial s^2$ vanish. A filled square labels the maximum of $\H$ and a filled circle a local minimum; both are stable fixed points. The orbits around them, illustrated by the banana-shaped dotted contour and the thin dotted circle, are referred to as libration and inner circulation orbits. They correspond to MMR capture and escape, respectively. The third fixed point, labeled by a filled diamond, is unstable; it is a saddle point.  Thick solid and thick dashed curves passing through it are the outer and inner separatrices. The thin dotted orbit outside the outer separatrix is an example of an external circulation orbit. Properties of these three kinds of orbits are listed below.
\begin{figure}
\includegraphics[width=0.98\linewidth]{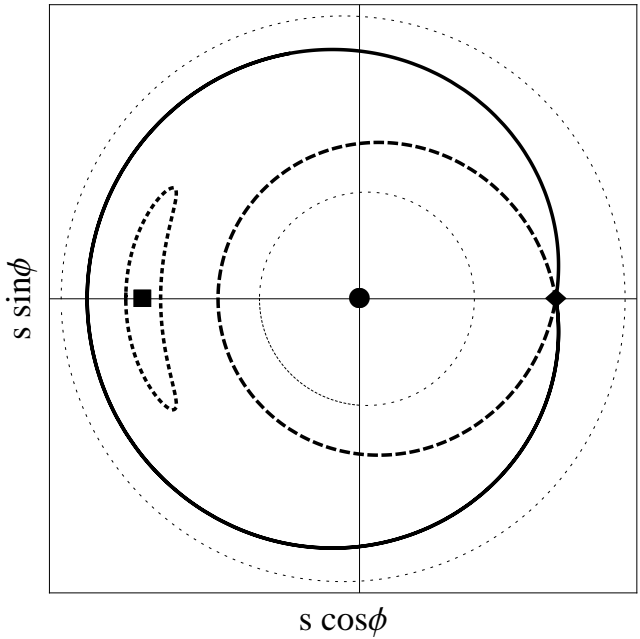}
\caption{\label{Contours-s2} Topology at $k>\kc$ for the $s^2$ MMR. The filled square, circle and diamond label respectively the local maximum of $\H$, the local minimum of $\H$, and the saddle point. The former two are stable fixed points, whereas the third is an unstable fixed point. Outer and inner separatrices are shown as thick solid and dashed curves passing through the unstable fixed point. The banana-shaped thick dotted curve is a libration orbit and corresponds to capture in MMR. The thin dotted circle inside the inner separatrix is an inner circulation orbit and corresponds to escape from MMR. The thin dotted circle outside the outer separatrix is an external circulation orbit.}
\end{figure}
\begin{enumerate}
\item External circulation: $n/n'>$ the exact resonant value. At conjunctions between Mimas and Tethys, the mean anomaly of Mimas circulates. Mimas is kicked by Tethys at every conjunction, but perturbations at different mean anomalies ultimately cancel. So the interaction is weak. Evolution under convergent migration may lead to either MMR capture or escape. At this stage, the system's destiny has not yet been determined. 
\item Libration: $n/n'>$ the exact resonant value. At conjunction with Tethys, the mean anomaly of Mimas librates.  Successive kicks occur at similar directions making the interaction strong. Libration orbits correspond to capture in MMR.
\item Internal circulation: $n/n'<$ the exact resonant value.  At conjunctions with Tethys, the mean anomaly of Mimas circulates.  As for external circulation, the interaction is weak.  The transition from external to internal circulation is equivalent to passage through resonance.  There is no path to return from internal circulation to libration.  
 \end{enumerate}

We express the rate at which dissipative processes alter $k$ by
\begin{equation}
{dk\over dt}={1\over\taun}-p{ s^2\over\taus}\, ,
\end{equation}
where $p$ is a numerical coefficient that depends on the dissipation mechanism and the mean motion ratio, $n/n'$. Timescales for convergent migration and inclination damping are denoted by $\taun$ and $\taus$. The former is dominated by Mimas's rate of tidal migration.\footnote{Mimas's orbit is interior to Tethys's, but Tethys is more massive.}  Noting that $(\dot n/n)_{\mathrm{tide}} \propto n^{13/3}$ \citep{1999ssd..book.....M} and the solar system's age is $5\Gyr$, currently $\taun\sim (13/3)\times 5\Gyr\sim 20\Gyr$. Cassini's laws indicate that Mimas's spin points almost parallel with its orbital angular momentum \citep{1969AJ.....74..483P}. Thus nodal precession is responsible for the time-varying part of tides in Mimas, yielding damping of  $s$ on timescale
\begin{eqnarray}
\taus&=&\left|s\over \dot s\right|_{\mathrm{tide}}\sim \left(19\over 5\right)^2\left(a\over R\right)^2{\mu\over \rho n^2 R^2}{Q\over\dot\Omega_J}\label{eq:taus}\\
&\sim& 5\times 10^3\Gyr\gg\taun\, ,
\end{eqnarray}
where we set $\mu\sim 4\times 10^{10}\dynpcmsq$ for the elastic rigidity of water ice. Mimas's ascending node precesses due to Saturn's oblateness at the rate $\dot\Omega_J\approx -n J_2(R_p/a)^2$, where $a$ is the orbital semi major axis of Mimas, $R$ its radius, $R_p$ the radius of Saturn, and $J_2\simeq 0.0163$ Saturn's quadrupole coefficient \citep{2004AJ....128..492J}. We set $\rho= 1.1\gpcc$ and $Q\simeq 100$ for Mimas \citep{1999ssd..book.....M}. Because $\tau_s\gg\tau_n$, $k$ keeps growing, and the inner and outer separatrices expand. The phase space areas they enclose, denoted by $\Ain$ and $\Aout$, are plotted by the solid and dashed curves in Fig.\refnew{A-k-s2}. The area between the inner and outer separatrices, $\Asep$, also enlarges with $k$, as plotted by the dot dashed curve in Fig.\refnew{A-k-s2}. At $k=\kc$, $\Ain=0$ and $\Aout=\Asep=\Ac\simeq 3.25\times 10^{-7}$. 
\begin{figure}
\includegraphics[width=0.99\linewidth]{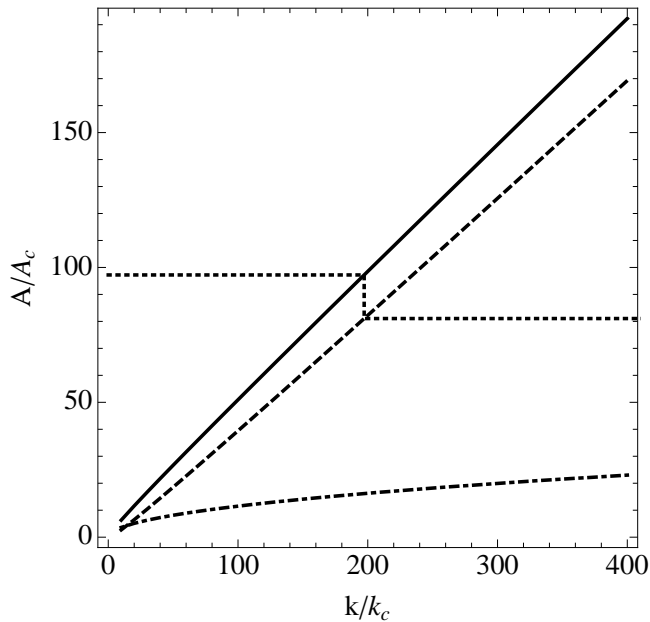}
\caption{\label{A-k-s2} Phase-space areas for the $s^2$ MMR: solid curve is $\Aout$, dashed is $\Ain$, and dot dashed is $\Asep$.  Dotted line segments depict the evolution of $A$ along an escape trajectory in the absence of inclination damping. Initially at $A_0$, $A$ jumps downward at $k=\kcap$, and subsequently remains constant at $A_i$. We estimate $\Ac\simeq 3.25\times 10^{-7}$, $\kcap\simeq 197\kc$, $A_0\simeq 3.164\times 10^{-5}$ corresponding to $I_0\simeq 0.364^\circ$, and $A_i\simeq 2.928\times 10^{-5}$ corresponding to $I_i\simeq 0.350^\circ$. At this $\kcap$, the capture probability is about $6.6\%$.}
\end{figure}

Here we describe evolution of Mimas along the dotted trajectory illustrated in Fig.\refnew{A-k-s2}.  
Let $I_0$ be the orbital inclination well before Mimas approaches the $s^2$ MMR.  The initial contour in phase space is a circle of radius $s_0\equiv \sin(I_0/2)$ centered at the origin. Absence of inclination damping makes the area ($A$) enclosed by the system's trajectory in phase space an adiabatic invariant; circulation around $A$ is fast on the tidal evolution timescale. For a while, $A$ stays at $A_0\equiv \pi s_0^2$.  Eventually the outer separatrix expands such that $\Aout=A_0$. Then Mimas passes through both the outer and inner separatrices and escapes the $s^2$ MMR.  The phase space area drops to $A_i=\Ain(\kcap)$ and remains there until the $ss'$ MMR is encountered.  The breaking of the adiabatic invariance of $A$ at separatrix crossing occurs because the libration period diverges there.  After Mimas passes the $s^2$ MMR, $A$ is an adiabatic invariant again.  Given Mimas's current state, we derive $A_i\simeq 2.92\times 10^{-5}$.\footnote{See the next subsection for an explanation.} Setting $\Ain(\kcap)=A_i$, we obtain $\kcap\simeq 197\kc$. Then $A_0=\Aout(\kcap)$ yields $A_0\simeq 3.164\times 10^{-5}$ and $I_0\simeq 0.364^\circ$.

Mimas could have been captured onto a libration orbit at the $s^2$ MMR.  For slow migration, capture probabilities depend solely on $\kcap$. Capture is certain if $\kcap<\kc$, or equivalently, if $A_0<\Ac=\Aout(\kc)$. The capture probability decreases with increasing $\kcap$ beyond $\kc$. It is about $6.6\%$ at $\kcap\simeq 197\kc$. These values are calculated from Equation (16) in \cite{1984CeMec..32..127B}. The basic idea is introduced below.

We denote the value of $\H$ at the separatrix by $\Hsep$. Libration orbits have $\H>\Hsep$, whereas internal circulation orbits have $\H<\Hsep$. After Mimas penetrates the outer separatrix, its trajectory initially hugs the outer separatrix from the inside and then transits to hug the inner separatrix from the outside. If it moves along the whole outer separatrix before it switches onto the inner separatrix, its $\H$ will end up larger than $\Hsep$, and thus capture is assured. Resonance escape will occur if 
the trajectory penetrates the outer separatrix just before it switches to the inner separatrix. Then
after the trajectory follows the entire inner separatrix, $\H$ will be smaller than $\Hsep$. Whether capture or escape occurs is determined by the phase ($\phi$) at which the trajectory penetrates the outer separatrix. 


\subsection{$ss'$ MMR}
After escaping the $s^2$ MMR with inclination $I_i$, Mimas approaches the $ss'$ MMR.  For this resonance, 
\begin{eqnarray}
k&=&12 s^2-{\kc^{3/2}\over 9 s}\cos\phi+{\dot\phi\over 2 n}\, ,\\
\H&=&k s^2-6 s^4+{2 \kc^{3/2}\over 9}s\cos\phi\, ,
\end{eqnarray}
where $\kc=-9f_{62}/2^{11/3} \mu' s'\simeq 5.3\times 10^{-6}$, $f_{62}\simeq -1.64$ and $\mu'\simeq 1.09\times 10^{-6}$ \citep{1999ssd..book.....M}. Currently, $I\simeq 1.57^\circ$, $I'\simeq 1.11^\circ$, $k\simeq 425\kc$, and $\phi$ librates around $0^\circ$ with semi amplitude $\phimax\simeq 97^\circ$ \citep{1992exaa.book.....S}. The topology of the phase space for $k>\kc$ is displayed in Fig.\refnew{fig:Contours}.
\begin{figure}
\includegraphics[width=0.99\linewidth]{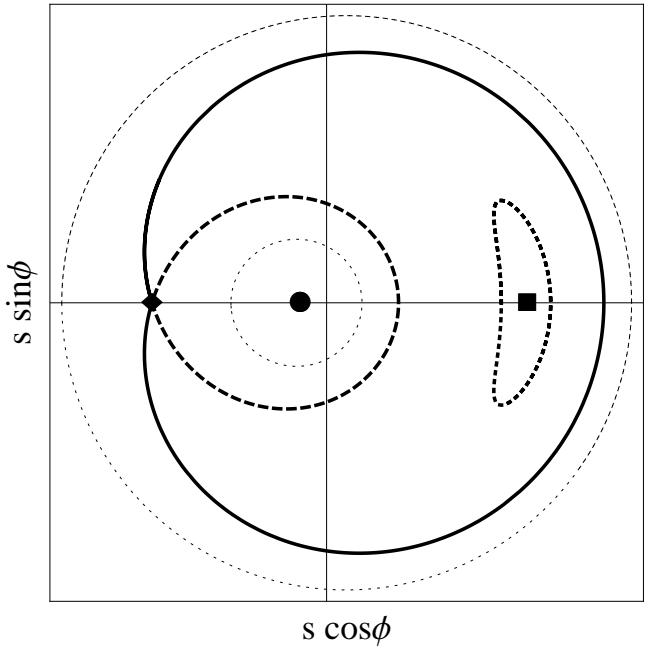}
\caption{\label{fig:Contours}Topology for the $ss'$ MMR with $k>\kc$. The filled square, circle, and diamond label respectively, the local maximum of $\H$, the local minimum of $\H$, and the saddle point. These are fixed points; the first and second are stable and the third is unstable. Solid thick and dashed thick curves going through the diamond symbol are the outer and inner separatrices. The banana-shaped thick dotted curve is a libration orbit, corresponding to capture in the MMR. The thin dotted circle inside the inner separatrix is an internal circulation orbit, corresponding to passage through the MMR. The thin dotted circle outside the outer separatrix is an external circulation orbit.}
\end{figure}
It is similar to that for the $s^2$ MMR, except that the local minimum of $\H$ (the filled circle in Fig.\ref{fig:Contours}) is located on the negative x-axis instead of the origin,  and the maximum (the filled square) is at $\phi=0^\circ$ rather than $\phi=180^\circ$. The classification of external circulation, libration and inner circulation orbits is the same as that described for the $s^2$ MMR. The concepts of $\Aout$, $\Ain$ and $\Asep$ also apply; all grow with $k$ (Fig.\ref{fig:A-k-ssp}).
\begin{figure}
\includegraphics[width=0.98\linewidth]{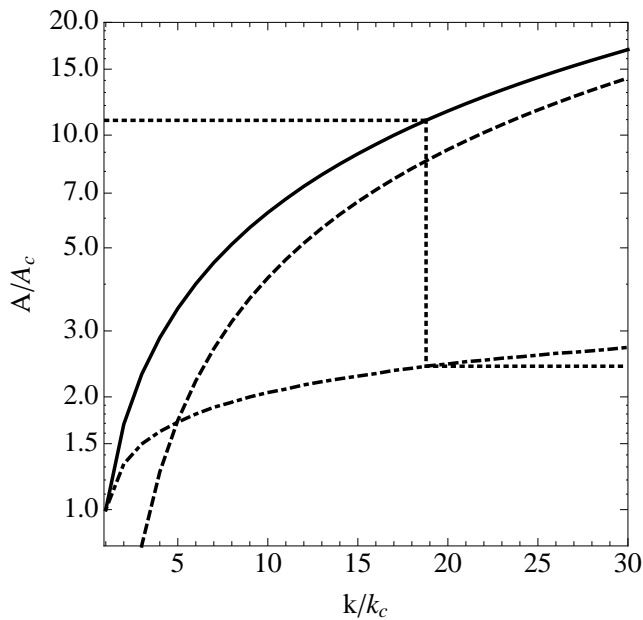}
\caption{\label{fig:A-k-ssp}  The solid, dashed, and dot dashed curves represent respectively, $\Aout$, $\Ain$, and $\Asep$. Dotted straight lines show the evolution of $A$ along a capture trajectory in the absence of inclination damping.}
\end{figure}

After Mimas passes through the $s^2$ MMR, its inclination angle is $I_i$, corresponding to a phase-space area $A_i=\pi\sin^2(I_i/2)=\pi s_i^2$. $A_i$ is shown by the upper horizontal dotted line in Fig.\refnew{fig:A-k-ssp}. The lower horizontal dotted line labels the area enclosed by Mimas's current orbit in phase space, $A_f\simeq 2.41\Ac\simeq 6.44\times 10^{-6}$. As Mimas moves deeper in the $ss'$ resonance, $k$ increases and the outer separatrix expands. At $k=\kcap$, $\Aout$ catches up with $A_i$. Mimas crosses the separatrix and gets caught on a libration orbit that marginally fits between the outer and inner separatrices. Therefore $\Asep(\kcap)=A_f$, leading to that $\kcap\simeq 18.8\kc$, and $A_i=\Aout(\kcap)\simeq 10.95\Ac\simeq 2.92\times 10^{-5}$. It follows that $I_i\simeq 2\arcsin((A_i/\pi)^{1/2})\simeq 0.350^\circ$. The corresponding capture probability is $6.0\%$ \citep{1984CeMec..32..127B}. It is a rare event.  Before arriving in its current state, Mimas avoided capture in the $s^2$ resonance, which also would have been a rare event. And if it had managed to pass through the $ss'$ resonance, it might still have been captured in the $s'^2$ resonance.  So finding Mimas in one of the three possible $4:2$ resonances with Dione is not so remarkable.  

Mimas's surface is heavily cratered \citep{2013A&A...553A..79D}. This raises an interesting question.  Could a large impact that occurred after Mimas was captured in the $ss'$ resonance have excited the observed libration of $\phi$? It took $\sim 424\kc/\dot k\sim 45\Myr$ for $k$ to grow from $\kc$ to its current value. The impact would have had to happen within this duration. Exciting a libration as large as $\phimax\sim 97^\circ$ requires the impact to impart a fraction of angular momentum comparable to $(\H(\phi=0)-\H(\phimax))^{1/2}\sim k_c^{3/4}(s(1-\cos\phimax))^{1/2}\sim 10^{-5}$.  With a minimum impact velocity comparable to the satellite's orbital velocity, this would require an impactor as large as $10\km$, which would produce a crater with diameter $\sim 170\km$ according to Equation (8) in \cite{2013A&A...553A..79D}. The largest crater on Mimas is $\sim 130\km$ across \citep{2013A&A...553A..79D} and it is implausible that it was produced during the past $45\Myr$.

Absence of inclination damping makes it safe for Mimas to keep evolving deeper into the MMR. Noting that $\dot k\simeq 1/\taun$ and $k\simeq 12 \sin^2(I/2)$ for $k\gg \kc$, Mimas's inclination will reach $16.6^\circ$ in another $5\Gyr$ about the time when the Sun will leave main-sequence.
%

\section{Titan-Hyperion}\label{sec:T-H}
For this resonance, the constants of motion, $k$ and $\H$ read
\begin{eqnarray}
k&=& 6 e'^2+{2^{1/2}\kc\over 3^2 e'}\cos\phi+{\dot\phi\over 4n'}\, ,\\
\H&=& ke'^2-3e'^4-{2^{3/2}\kc\over 3^2}e'\cos\phi\, ,
\end{eqnarray}
where $e'$ is the eccentricity of Hyperion. The critical value for $k$ is 
\begin{equation}
\kc={3^{4/3}\over 2^{5/3}}(f_{31}\mu)^{2/3}\simeq 0.0115\, ,
\end{equation}
where $f_{31}\simeq 0.825$, and $\mu\simeq 2.37\times 10^{-4}$ is the mass ratio of Titan over Saturn \citep{1999ssd..book.....M}. Currently $e'\simeq 0.104$, $k\simeq 5.5\kc$, and $\phi$ librates around $180^\circ$ with semi-amplitude $\phimax\simeq 36^\circ$ \citep{1992exaa.book.....S}. The topology of the phase space is the same as that of the $ss'$ MMR for Mimas-Tethys (Fig.\ref{fig:Contours}) except that the maximum $\H$ (the filled square) is located at $\phi=180^\circ$ and the local minimum and unstable fixed point are at $\phi=0^\circ$.  The phase-space area $A_f\simeq 0.092\Ac<\Ac$ which if conserved since capture implies a capture probability of $100\%$. If the system never penetrated the outer separatrix, its  current phase-space area is the same as it was prior to capture, i.e., $A_i=A_f$ implying an initial eccentricity for Hyperion
\begin{equation}
e'_i\simeq \sqrt{A_i/\pi}\simeq 0.019\, .
\end{equation}
Hyperion is so far away from Saturn that the tidal dissipation of $e'$ is negligible; we estimate $\tau_{e'}\sim 6\times 10^7\Gyr$, by replacing $\dot\Omega_J$ in Eq.\refnew{eq:taus} by $n$ and Mimas's parameters by Hyperion's. Hyperion's irregular shape indicates it might have suffered a huge impact, which if directed randomly would induce comparable eccentricity and inclination. Its current inclination of $\sim 7.5\times 10^{-3}$ radian is much smaller than its current eccentricity but similar to $e'_i$, making impact before capture a plausible explanation for Hyperion's initial eccentricity. Frequent impacts likely accompanied the formation of satellites and the Titan-Hyperion MMR probably dates back to this time.  

In what follows, we discuss the mechanism of convergent migration which leads to MMR capture. Both tidal and disk induced migrations act more efficiently on Titan, the more massive and less distant satellite. Consequently, convergent migration suggests that Titan moved outward. However, currently its tidal migration rate is slow. We find $\taun\sim 1189\Gyr$, utilizing $\taun\sim 20\Gyr$ for Mimas and $\taun\propto a^{13/2}/m_s$ \citep{1999ssd..book.....M}, where $m_s$ is the mass of the satellite. Saturn was more luminous and larger \citep{1992exaa.book.....S} when it was young, during which stage Titan may migrated significantly outward.  Alternatively,  Titan's outward migration might have been due to interactions with the disk from which the satellites formed.  Although theoretical considerations suggest that special conditions are required to produce outward migration \citep{2013arXiv1312.4293B}, they are likely to have existed, as implied by observations of transiting planets at large separations \citep{2013Sci...340..572H}. 


\section{Conclusion}\label{sec:conclusion}

The satellite pairs of Saturn, Mimas-Tethys and Titan-Hyperion, are special in their depth in resonance ($k\gg\kc$) and the large librations of their resonance arguments. Large depth in resonance is only safe in absence of inclination or eccentricity damping. Otherwise, escape from resonance would occur on the relevant damping timescale \citep{2014AJ....147...32G}. Other resonant pairs survive either because they remain in shallow resonance ($k\ll \kc$) or because they only venture deep in resonance for times short compared to these damping timescales.

Convergent migration probably was responsible for the formation of the MMRs between Mimas-Tethys and Titan-Hyperion.  For the former, it is mainly due to Mimas's tidal migration. For the latter, although the current tidal interaction between Saturn and Titan is too slow, it may have been adequate when Saturn was much bigger shortly after its birth. Disk induced migration is an alternative. 

Current librations of the resonance arguments, allow us to deduce the initial eccentricity of Mimas and the initial eccentricity of Hyperion prior to capture in MMRs.  We find an initial free inclination of Mimas ($I_0\simeq 0.36^\circ$) and and initial free eccentricity of Hyperion ($e'_i\simeq 0.019$). The former implies a capture probability into $ss'$ resonance $\sim 6\%$, whereas the latter guarantees capture in the $4:3$ resonance. Our estimate of the capture probability for Mimas-Tethys agrees roughly with $4\%$ given by \cite{1972MNRAS.160..169S}.  \cite{1972MNRAS.160..169S} also explored whether close encounters between Titan and Hyperion can account for their current libration. Conjunction with Titan at Hyperion's pericenter might supply a strong enough perturbation, but, as the author pointed out, the $4:3$ MMR constrains conjunctions to take place close to Hyperion's apocenter.

\section*{acknowledgement}
I thank Peter Goldreich for his inspiration, suggestions and comments.

\end{document}